\documentclass[twocolumn,prl,showpacs,preprintnumbers,amsmath,amssymb,superscriptaddress,unsortedaddress]{revtex4}
\usepackage{graphicx}% Include figure files
\usepackage{dcolumn}% Align table columns on decimal point
\usepackage{bm}% bold math
\usepackage{epsf,epsfig,latexsym}

\begin{document}

\title{High $m_T$ pion and proton correlations in central PbPb collisions at $\sqrt{s_{NN}}=17$GeV}

\newcommand{\nbi} {Niels Bohr Institute, University of Copenhagen, Denmark}
\newcommand{\lanl}{Los Alamos National Laboratory, Los Alamos, NM 87545}
\newcommand{\Nantes}{Nuclear Physics Laboratory of Nantes, 44072 Nantes, France}
\newcommand{\Columbia}{Department of Physics, Columbia University, New York, NY 10027}
\newcommand{\Hiroshima}{Hiroshima University, Higashi-Hiroshima 739-8526, Japan}
\newcommand{\CERN}{CERN, CH-1211 Geneva 23, Switzerland}
\newcommand{\Zagreb}{Rudjer Bo\v{s}kovi\'c Institute, Zagreb, Croatia}
\newcommand{\Ohio}{Department of Physics, The Ohio State University, Columbus, OH 43210}
\newcommand{\SUNY}{State University of New York, Stony Brook, NY 11794}
\newcommand{\Lund}{Department of Physics, University of Lund, S-22362 Lund, Sweden}
\newcommand{\Texas} {Texas A\&M University, College Station, Texas 77843-3366}
\newcommand{\BNL} {Brookhaven National Laboratory, Upton, New York 11973}
\author{I.G.~Bearden}\affiliation{\nbi}
\author{H.~B\o ggild}\affiliation{\nbi}
\author{J.~Boissevain}\affiliation{\lanl}
\author{P.H.L.~Christiansen}\affiliation{\nbi}
\author{L.~Conin}\affiliation{\Nantes}
\author{J.~Dodd}\affiliation{\Columbia}
\author{B.~Erazmus}\affiliation{\Nantes}
\author{S.~Esumi}\affiliation{\Hiroshima}
\author{C.W.~Fabjan}\affiliation{\CERN}
\author{D.~Ferenc}\affiliation{\Zagreb}
\author{A.~Franz}\affiliation{\CERN}
\author{J.J.~Gaardh\o je}\affiliation{\nbi}
\author{A.G.~Hansen}\affiliation{\nbi}
\author{O.~Hansen}\affiliation{\nbi}
\author{D.~Hardtke}\affiliation{\Ohio}
\author{H. van~Hecke}\affiliation{\lanl}
\author{E.B.~Holzer}\affiliation{\CERN}
\author{T.J.~Humanic}\affiliation{\Ohio}
\author{P.~Hummel}\affiliation{\CERN}
\author{B.V.~Jacak}\affiliation{\SUNY}
\author{K.~Kaimi} \altaffiliation{Deceased} \affiliation{\Hiroshima}
\author{M.~Kaneta}\affiliation{\Hiroshima}
\author{T.~Kohama}\affiliation{\Hiroshima}
\author{M.~Kopytine}\affiliation{\SUNY}
\author{M.~Leltchouk}\affiliation{\Columbia}
\author{A. Ljubi\v{c}i\'c Jr.}\affiliation{\Zagreb}
\author{B.~L\"orstad}\affiliation{\Lund}
\author{N.~Maeda}\affiliation{\Hiroshima}
\author{L.~Martin}\affiliation{\Nantes}
\author{A.~Medvedev}\affiliation{\Columbia}
\author{M.~Murray}\affiliation{\Texas}
\author{H.~Ohnishi}\affiliation{\Hiroshima}
\author{G.~Pai\'c}\affiliation{\CERN}
\author{S.U.~Pandey}\affiliation{\Ohio}
\author{F.~Piuz}\affiliation{\CERN}
\author{J.~Pluta}\affiliation{\Nantes}
\author{V.~Polychronakos}\affiliation{\BNL}
\author{M.~Potekhin}\affiliation{\Columbia}
\author{G.~Poulard}\affiliation{\CERN}
\author{D.~Reichhold}\affiliation{\Ohio}
\author{A.~Sakaguchi}\affiliation{\Hiroshima}
\author{J.~Schmidt-S\o rensen}\affiliation{\Lund}
\author{J.~Simon-Gillo}\affiliation{\lanl}
\author{W.~Sondheim}\affiliation{\lanl}
\author{T.~Sugitate}\affiliation{\Hiroshima}
\author{J.P.~Sullivan}\affiliation{\lanl}
\author{Y.~Sumi}\affiliation{\Hiroshima}
\author{W.J.~Willis}\affiliation{\Columbia}
\author{K.~Wolf} \altaffiliation{Deceased}\affiliation{\Texas}
\author{N.~Xu}\affiliation{\lanl}
\author{D.S. Zachary}\affiliation{\Ohio}

\collaboration{The NA44 Collaboration, \today}\noaffiliation

\begin{abstract}
For central  PbPb
collisions at $\sqrt{S_{nn}}=17.3$ GeV
we have made the first two-dimensional measurement of the pp
correlation  function.
These data
extend the range of previous studies of
HBT radii by
a factor of two in $m_T$. %
 They % data, and all our previous HBT results, 
are consistent with a hydrodynamic
 interpretation and
microscopic models that include hadronic rescattering and transverse expansion.     
We also report new data on pion correlations. 
%We have also measured 
The two particle correlations of  negative pions
 at $m_{T} = 0.92$ GeV imply source % We find 
radii that are smaller than typical hydrodynamic fits and transport model simulations.
 It is possible that these fast
pions may have left the source before the build up of hydrodynamic flow.
\end{abstract}
\pacs{25.75.Gz, 25.75.Ld}
\keywords{two-proton correlations; two pion correlations,  nucleus-nucleus; heavy-ion
collisions; proton source size; pion source size CERN SPS}

\maketitle

%% \section{Introduction}
Experimental studies of high-energy nuclear collisions at the AGS,
SPS and RHIC accelerators %(with $\sqrt{S_{nn}}$ from  2 to 200 GeV)
(with $\sqrt{S_{nn}} =2 \rightarrow $200 GeV)
have revealed many interesting features of hot and dense nuclear matter,
and some characteristic signatures of a quark-gluon-plasma phase
have been reported~\cite{QM}. 
If such a high density state were formed in the initial stages of the 
reaction the high initial presure would result in a significant transverse flow of 
the hadrons that originate from the participant zone. Depending on the time
scale and the amount of rescattering the system might be expected to be in
local thermodynamical equilibrium. This conjecture is supported by the
 linear increase of the single
particle inverse slopes with mass~\cite{PRL78,NA44:dt}.
 
The Hanbury-Brown Twiss, (HBT), technique uses the interference
of particle wavefunctions to infer the angular size of stars or the length
scales of  nuclear systems from the two particle correlation function \cite{HBTGGLP}.
The ``radii" measured by  HBT   can be thought of as
``lengths of homogeneity'' of the source %that reflect the  lengths scales of the
which depend upon
velocity and/or temperature
 gradients  \cite{SINYUKOV,CSOR94A,WEID99A}.
Heavy Ion HBT measurements over a wide range of energies
  \cite{Ahle01,NA35,NA44,NA49,WA98,PRC(John),NA44KK,StarHBT, PhenixHBT}
 show a drop of radii with $m_T$ $ \equiv \sqrt{p_T^2+ m^2}$
consistent with %hydrodynamics 
a hydrodynamical interpretation 
\cite{Akkelin95,Wiedemann96,Lorstad96,HeinzJacak}.

Faster particles are more likely to come from initial
 collisions of the incoming baryons, so  at some momentum 
 we might expect the hydrodynamic
 approach to break down. This may have already been seen at RHIC 
($\sqrt{s_{NN}}=200$GeV) where
  the strength of elliptic flow falls below the hydrodynamic prediction 
for $p_T > 2$GeV/{\it c}
 \cite{Adler:2002ct}. One might expect this %point
 to occur at lower $p_T$ at
 $\sqrt{s_{NN}}=17$GeV
 since the multiplicity is lower.

 NA44 is a focusing spectrometer \cite{PRC(John)}. The dipole and quadrupole magnets produce a magnified image of
 the target in the tracking detectors. This has the advantage that particles that are close in
 momentum are not necessarily close in position. Since all 
tracking is done after the magnets
 we only have to reconstruct straight lines.
 A  high resolution pad  chamber  sits in the focal plane and
 gives the magnitude of the momentum.
 Downstream of the focus
 the direction of tracks is
 measured by strip chambers
 and scintillating hodoscopes.
The momentum resolution, including effects of multiple scattering
in the target, is $\approx$ 11MeV/{\it c}  for $p_x$, $p_y$, and $p_z$.

A beam counter
selected events for which single lead
ions hit the 3.8mm lead target.
 Behind the target, two scintillator bars were used to trigger on high multiplicity events.
% Off line
 An annular  silicon detector
 with  512 pads which was 
 not part of the trigger measured  $dN^\pm/d\eta$    ($\eta = \ln \tan{\theta/2})$
in the
pseudorapidity range $ 1.5 \le \eta \le 3.3$. % \cite{NA44Mult}.
We used  events in the top 18\% of the multiplicity distribution
as we did for our earlier $PbPb$
results  \cite{PRC(John),NA44KK}.
For all   data sets except the lower $m_T$ protons
the spectrometer was set to
accept particles of momentum 6-9 GeV/{\it c} and % sat at
was positioned at 
131 mrad with respect to the beam axis.
 The setting gave a  $p_T$ window of 0.7--1.4 GeV/{\it c}.
 For the lower $m_T$ proton sample
 the momentum window was 5.2--8.0GeV/{\it c} and the spectrometer
 angle was 44 mrad. The resulting $p_T$ range was
0.18--0.50GeV/{\it c}.
 The rapidity range  of the data was 2.4--2.9 and the system is symmetric about
 y=2.9 the center of mass rapidity. % Plots of 
The  $p_T$ and y acceptance can be found
in  \cite{Masashi}.

Particles were  identified  by combining time of flight measurements
from the
hodoscopes
with velocity information from
three threshold-type gas Cherenkov counters (C1, C2, TIC).
 C2 was set to fire only on pions.
For the higher momentum proton data C1 was
used to reject kaons. For the lower $m_T$ proton data   kaons
were rejected exclusively by their flight time.
For the $\pi^-$ sample the TIC %Track Imaging Cerenkov, 
was used to confirm that
each track was a pion \cite{TIC}.
Contamination of 
the samples by other particles was less than 2\%.
However there is significant feed-down of hyperons %$\Lambda$s and $\Sigma_0$s
into the proton sample, ranging from 28\% for the lower $m_T$ sample to
13\% for $\langle m_T \rangle $ = 1.45 GeV \cite{Masashi}. This tends to
smear the  correlation function and was accounted for
by including a fraction of ``fake" protons in the Monte Carlo \cite{NA44pp}. The spectrum
of hyperons was taken from RQMD with the yield scaled to match
 NA49 and WA97 results \cite{NA49:lambda,Ant00}.

%% \section{Analysis}
The two-particle correlation function is defined by 
\begin{equation} 
C_2(\vec{p}_1,\vec{p}_2)= \frac{P_2(\vec{p}_1,\vec{p}_2)}
        { P_1(\vec{p}_1)P_1(\vec{p}_2)}
        \approx 
	\frac {Real(\vec{p}_1,\vec{p}_2)}
         {Back(\vec{p}_1,\vec{p}_2)}, 
\label{eq:C_raw}
\end{equation} 
where the numerator is the joint probability of detecting two particles  
with momenta $\vec{p}_1$ and $\vec{p}_2$, 
while the denominator is the product of the probabilities of detecting 
the single particles. % with momenta $\vec{p}_1$ and $\vec{p}_2$.
The denominator was obtained by mixing tracks  from
two randomly selected different events.
%Ten mixed background pairs
%were generated for each real pair to reduce statistical uncertainties
%from the background.
For protons, the attractive effect of the strong interaction
competes with the negative correlation due to Fermi-Dirac statistics and Coulomb
repulsion.
This % combination 
results in  characteristic ``dip-peak'' structure
in the two-proton correlation function.
The height of the correlation 
peak decreases as the source size increases
\cite{Koonin,Lednicky,Pratt,Ghisal}. % Should I reference these here.

The true correlation function is distorted by the momentum resolution, 
Coulomb repulsion
and   by residual two particle correlations in the background.
The magnitude of these effects depends on the size of the source.
For the pions we 
we evaluated  them using
an iterative  Monte Carlo (MC) simulation.
%based iteration method.
A trial correlation function was
used to generate events that we tracked through  the spectrometer and
 analyzed in exactly the same way as the data.
The Coulomb wave %integration
method
was employed to simulate Coulomb effects ~\cite{Pratt86}.
This method was used to derive correction factors to
the correlation function and the process was repeated until  
the changes in the corrections from step to step were insignificant ~\cite{NA44,zajc84}.
The theoretical correlation function for the pp data was generated by selecting 
 proton pairs from a given source model, calculating the weight due to quantum 
 statistics and final state interactions (strong and Coulomb) and propagating the 
 particles through the Monte Caro.  
 We searched for a source distribution that produced the
minimum $\chi^2$ with the data after it was fed through the Monte Carlo \cite{NA44pp}.

%% \section{Result and discussion}
$C_2(\vec{p}_1,\vec{p}_2)$ is a function of 6 variables;
3 for  the
total momentum $\vec{P}=\vec{p}_1+\vec{p}_2$, and
3 for the momentum difference  $\vec{Q}=\vec{p}_1-\vec{p}_2$.
We decompose $\vec{Q}$ into
$Q_L$, the projection of the momentum difference onto the beam axis,
$Q_{out}$  which is parallel to
$\vec{P}$ and $Q_{side}$ which is perpendicular to $Q_{out}$ and $Q_L$.
The data are analysed in  the Longitudinal Co-Moving System in which
$p_{z1}=- p_{z2}$.

To measure correlation functions efficiently %is that we trigger
we trigger
on events with pairs of particles that
have a small momentum difference in
one direction while allowing large momentum differences in the other
two dimensions.
In our ``vertical setting" the quadrupoles produce 
 a small
 momentum
acceptance in the horizontal  space $p_x$, and  a wide
 acceptance in the vertical direction $p_y$.
 Since the spectrometer lies in the horizontal plane this
 setting allows only a small range in $Q_{side}$ but a large
 range in $Q_{out}$.
 In the ``horizontal setting" %the quadrupoles produce 
we have 
a wide momentum
acceptance in
$p_x$ and a small
 acceptance in $p_y$.   This setting allows only a small range in $Q_{side}$ but a large
 range in $Q_{out}$. Both settings have a large acceptance in $Q_{Long}$.
 The lower $m_T$ proton  data were taken in the vertical setting while all
 other data were taken in the horizontal setting.

\begin{figure}%[b]
\label{hbt}
\epsfig{file=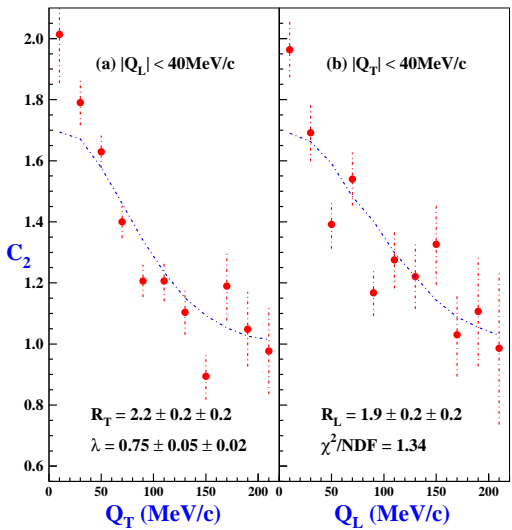,width=7cm}
\caption{The  $\pi^{-}$ correlation function for 
$\langle m_T \rangle = 0.92GeV$ projected onto %(a) the $Q_T$ axis and (b) $Q_L$ axis. 
 the (a) $Q_T$   and  (b) the $Q_L$axies. 
Only statistical errors are shown.
 The  lines are projections of the fit.}
\label{fig:c2hipi}
\end{figure}

For pions we used the maximum likelihood method to fit C2 with
\begin{equation}
C_2(Q_T,Q_L)=D(1+ \lambda e^{-Q_T^2R_T^2-Q_L^2R_L^2} ),
\label{eq:C_2_2dim}
\end{equation}
where 
$R_T$ and $R_L$  parametrize the
size of the source in the transverse and beam directions and $D$ is a free
parameter used for normalization.  
The $\lambda$ factor is a measure of the chaoticity of quantum states of the source;
the fraction of pions from resonances and experimental factors that decrease the correlation function. 
For these data $Q_T \approx Q_{out}$.
Figure~\ref{fig:c2hipi} shows projections of the correlation function  and the fit
onto the $Q_T$ and $Q_L$ axis.
The $\lambda$ parameter is larger than for our lower $p_T$ pion measurements. This is expected since
the effect of resonances, which tend to wash out $C_2$ and reduce  $\lambda$, should be less important
as $p_T$ increases.
 The systematic
uncertainties in the fit parameters reflect the effect of
(i) cut parameters to define a track,
(ii) cut parameters to select pairs, 
(iii) momentum resolution, 
(iv) two-track resolution, 
(v) momentum distribution of particle production in MC, and
(vi) fitting to finite bins. 

\begin{figure}[floatfix]%[b]
\epsfig{file=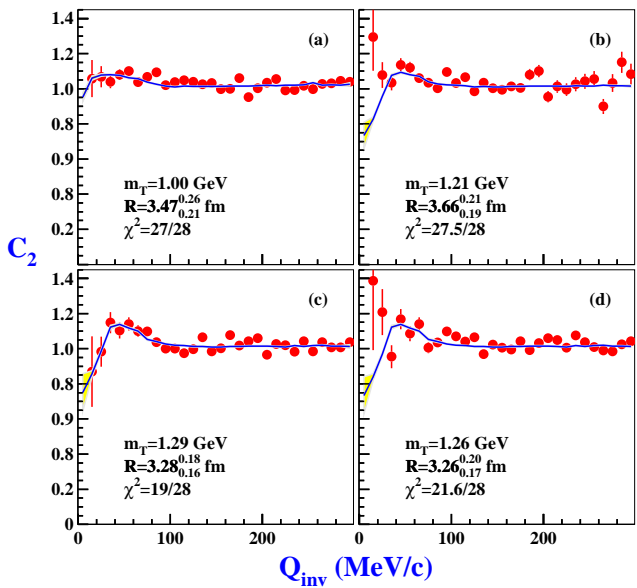,width=8.cm}
\caption{Projections of the $pp$ correlation function onto the $Q_{inv}$ axis
for samples with different  $\langle m_T \rangle$.
Only statistical errors are shown.
 The  lines show the projections of the fit.}
\label{fig:c2pp}
\end{figure}

Figure~\ref{fig:c2pp} shows 
the $pp$ correlation function as a function of 
$Q_{inv} \equiv \sqrt{Q^2 - \Delta E^2}$ %axis 
for four bins in  $\langle m_T \rangle$.
At $\langle m_T \rangle=1$ GeV our result is consistent with NA49 \cite{NA49pp}. 
$R_{inv}$ increases from $pPb$ to $SPb$ and finally to $PbPb$, \cite{NA44pp}.
We have also analysed the pp correlation in two dimensions to extract $R_T$ and $R_L$.
%There is a negative correlation between the errors on  $R_T$ and $R_L$.
Table \ref{tab:radii}  summarizes
our data and compares them to two models.

RQMD 2.04  is a string model which includes reinteractions between produced hadrons \cite{RQMD}.
It gives a good description of the proton data, as it does for our low $m_T$ pion 
results \cite{PRC(John),Sullivan:vb}.  The decrease of the radii with $m_T$ has already been
noticed in the model for pions and results from correlations between position and momentum caused
by flow \cite{Fields}. This decrease is more rapid for $R_L$ than for $R_T$ since the longitudinal
flow is stronger.
Nexus 2.00 uses pertubative QCD to treat  hard  nucleon-nucleon interactions and pomeron
exchange for soft interactions \cite{Nexus}. It produces proton radii that are too small. 
Turning on hadronic reinteractions and resonance production 
results in larger radii closer to the data. % \cite{Nexus2b}.
RQMD fails to reproduce both
pion radii at $m_T$ = 0.92GeV.  Nexus reproduces $R_L$ but overestimates $R_T$ by
a  factor of two.
\begin{table}   
\caption{\label{tab:radii} PbPb source radii versus $m_T$ from data and  models.}
\begin{ruledtabular}
%\centering
  \begin{tabular}{|c|c|c|c|c|c|c|c|}
%    \hline
       &   \protect{$m_T$}   & \multicolumn{3}{c|}{ $R_T$ (fm)}
                                           & \multicolumn{3}{c|}{ $R_L$ (fm)} \\   %\cline{3-8}
   & (GeV) & Data & Rqmd & Nexus & Data & Rqmd & Nexus  \\   \hline
    p   &  1.00  & $3.6_{-0.6}^{+0.7}$   & 3.7 & 3.1 & $3.5_{-0.9}^{+1.2}$    &  4.6 & 1.7 \\
           &  1.21  & $4.0_{-0.3}^{+0.4}$   & 3.6 & 3.2 & $3.5_{-0.5}^{+0.6}$    &  3.8 & 1.9 \\
          &  1.29  & $3.2_{-0.2}^{+0.2}$   & 3.5 & 3.2 & $3.3_{-0.4}^{+0.6}$    &  3.5 & 2.0 \\
          &  1.43  & $3.4_{-0.2}^{+0.3}$   & 3.4 & 2.8 & $2.9_{-0.3}^{+0.4}$    &  3.1 & 2.0 \\  \hline
$\pi^-$ & 0.92  & $2.2 \pm 0.3 $   & 3.9 & 4.5 & $1.9 \pm 0.3 $    &  3.1 & 2.0 \\   %\hline
  \end{tabular}
\end{ruledtabular}
\end{table}

 Under certain conditions ~\cite{HeinzJacak}
one can derive analytic expressions for the radii:
%\begin{eqnarray}
\begin{equation}
\frac{1}{R_T^2}  \approx  \frac{1}{R^2}   \left( 1+ \eta_f^2 \frac{m_T}{T} \right);
\label{eqn:RtvMt}
\end{equation}
\begin{equation}
\frac{1}{R_L^2}  \approx    \frac{1}{\tau_0^2 \Delta \eta^2}  \left( 1 + \Delta \eta^2 \frac{m_T}{T} \right);
\label{eqn:RLvMt}
\end{equation}
where $R$ is the geometrical size of the source,
$\tau_0$  and $T_0$ are the  time and temperature of freeze-out, $ \delta \eta$ represents the
width in rapidity  and $\eta_f$ is the transverse rapidity. The radii also 
depend on the particle's mass but this effect is very small for reasonable values of $\eta_f$.
Since $\eta_f< \Delta \eta$, $R_L$ should drop more rapidly with $m_T$ than $R_T$ 
because the longitudinal flow is stronger than the
transverse flow.

\begin{figure}%[tbph]
\epsfig{file=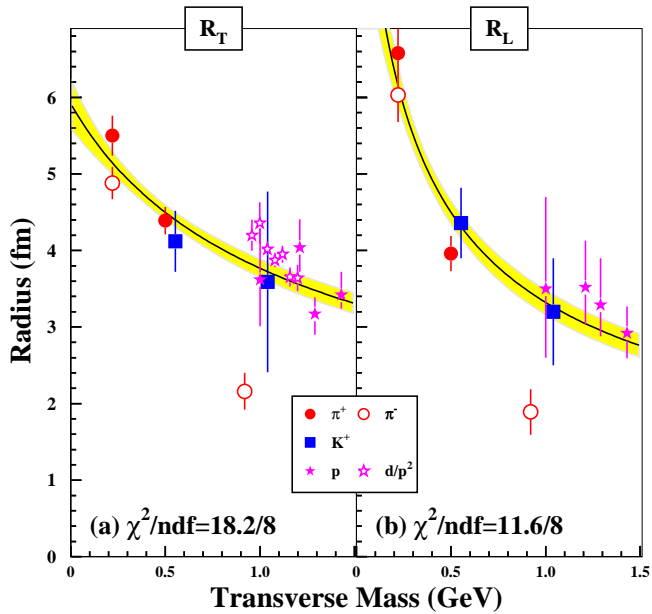,width=9.5cm}
\caption{NA44 pion, kaon and proton source sizes versus $m_T$
~\cite{PRC(John),NA44KK,Murraydpp}. Both radii are fit to the form $1/R^2=p0+p1\cdot M_T$.
The lines and bands show the results and errors of the fits.}
\label{fig:rmtna44}
\end{figure}

Figure~\ref{fig:rmtna44}
shows the $m_T$  dependence of $R_T$ and $R_L$ for all NA44 correlation
 and coalescence measurements ~\cite{PRC(John),NA44KK,Murraydpp}.
The proton coalescence analysis gives only one radius
which is mainly sensitive 
to the transverse size, %( see discussion in Sec.~3 of \cite{Murraydpp} and
(see Eqn.~6.3 of \cite{SCHEIBL}). These radii are in good agrement with the pp correlation data.
   $R_L$  decreases faster with $m_T$ than does $R_T$ as expected from    RQMD  simulations  and
   Eqns~\ref{eqn:RtvMt} and \ref{eqn:RLvMt}. The lines in Fig.~\ref{fig:rmtna44} 
are the  result of fitting our systematics to  the form $1/R^2=p0+p1\cdot M_T$.
The bands show the errors on the fits.   The
   $\chi^2/{\rm ndf}$ is poor because the
$\pi^-$ radii at $\langle m_T \rangle\approx 0.92$ GeV  are
 smaller than the trend expected from our other data. Removing the high $m_T$ 
 $\pi^-$ radii from the fit 
 gives $\chi^2/{\rm ndf} \approx 1$ but has a negligible effect on the fit parameters.

At $m_T = 0.22$GeV the    negative pions have source radii that are 9-11\%
smaller than for the positive pions. However the $\pi^-$ and $\pi^+$
correlation functions themselves are consistent
and   non Gaussian in shape \cite{PRC(John)}.
 RQMD
simulations suggest that this is due to resonance decays, particularly of the
$\omega$. Resonances tend to reduce the $\lambda$ parameter when fitting $C_2$ to
Eqn.~\ref{eq:C_2_2dim}. This effect decreases
rapidly with $p_T$ \cite{Fields}. It is possible that interaction with the residual nuclear
charge could reduce the $\pi^-$ radius as compared to the $\pi^+$ radius.
Again %one might expect that
this effect should %would
be stronger at low $p_T$.

In summary, we have made the first two-dimensional measurement of the pp correlation function
in relativistic nucleus-nucleus collisions.
NA44 has measured HBT radii over a significantly larger  
 $m_T$ range than any
other experiment.
The pp data, and all our previous HBT results, are consistent with a hydrodynamical interpretation and
microscopic models that include hadronic rescattering and transverse expansion.
$R_L$ drops more rapidly with $m_T$ than $R_T$ because the longitudinal flow is stronger than the
transverse flow.
At $m_T$ = 0.92GeV we find that the $\pi^-$ radii
are somewhat smaller than  expected from the
trend of our other data and are not easily explained by either hydrodynamic fits or the RQMD model.
It may be that  these fast pions  left
the source  before the buildup of flow and so offer  a glimpse of the hadronic system at an early point
in its expansion.

We %wish to
thank the staff of the CERN %PS-SPS
accelerators % complex
for their excellent work.
We are  grateful for financial support from
the Danish Natural Science Research Council;
the Japanese Society for the Promotion of Science
and the Ministry of Education, Science and Culture;
the Swedish Science Research Council;
the Austrian Fond f\"{u}r F\"{o}rderung der Wissenschaftlichen Forschung; 
the US National Science Foundation and the Department of Energy.


\begin{thebibliography}{99}
\bibitem{QM} {\it Proceedings of Quark Matter 2001},
T.J. Hallman, D.E. Kharzeev, J.T. Mitchell and T. Ullrich , [Nucl. Phys. {\bf A698},
 1c-676c (2002) and Quark Matter 2002  http://alice-france.in2p3.fr/qm2002/ .
\bibitem{PRL78} I.~G.~Bearden {\it et al.}, Phys. Rev. Lett. {\bf 78}, 2080 (1997).
%\cite{NA44:dt}
\bibitem{NA44:dt} I.~G.~Bearden {\it et al.},
%``Deuteron And Triton Production With High Energy Sulphur And Lead Beams,''
Eur.\ Phys.\ J.\ C {\bf 23}, 237 (2002).
%%CITATION = EPHJA,C23,237;%%

\bibitem{HBTGGLP} R.~Hanbury-Brown, R.~Q. Twiss, Nature {\bf 178} 1046 (1956) and
G.~Goldharber, S.~Goldharber, W.~Lee, A.~Pais Phys. Rev. {\bf 120} 300 (1960).
%Making such historical references it is necesasary to mention the paper where 
% for the first time the space/time size (not angular size) measurements for 
% particle physics were proposed. It is: 
% G.I. Kopylov, M.~I. Podgoretsky, Sov. J. Nucl.Phys. 15, 219 (1972), 
% G.I. Kopylov, Phys. Lett. 50, 472 (1974) 
\bibitem{SINYUKOV} Yu.~M. Sinyukov, Nucl. Phys. {\bf A566} 589c (1994).
\bibitem{CSOR94A} T. Cs{\"o}rg\H o and B. L{\"o}rstad,  Phys. Rev. {\bf C} {\bf 54} 1390 (1996).
\bibitem{WEID99A} U.~A. Weidemann and U. Heinz, Phys. Rep {\bf 319} 145 (1999). %  nucl-th/9901094.

\bibitem{Ahle01} L. Ahle, {\it et al.}, Phys. Rev. {\bf C66} 054906 (2002).
%`` System, centrality, and transverse mass dependence of two-pion correlat% ion radii in heavy ion collisions at 11.6 and 14.6 A-GeV"
%nucl-ex/0204001 %The E802 Collaboration
\bibitem{NA35}
A.~Bamberger {\it et al.}, Phys. Lett. {\bf B203}, 320 (1988);
T.~Alber {\it et al.}, Z. Phys. {\bf C66}, 77 (1995).
\bibitem{NA44}
H.~B\o ggild {\it et al.}, Phys. Lett. {\bf B302}, 510 (1993).
%; Phys. Lett. {\bf B349}, 386 (1995);
%H.~Beker {\it et al.}, Z. Phys. {\bf C64}, 209 (1994); 
%Phys. Rev. Lett. {\bf 74}, 3340 (1995);
%K.~Kaimi {\it et al.}, Z. Phys. {\bf C75}, 619 (1997).
\bibitem{NA49}  H.~Appelsh\"{a}user {\it et al.}, Eur. Phys. J. {\bf C2}, 661 (1998).
\bibitem{WA98}  M.~M.~Aggarwal {\it et al.}, Eur. Phys. J. {\bf C16}, 445 (2000).
\bibitem{PRC(John)}  I.~G.~Bearden {\it et al.},
Phys. Rev. {\bf C58}, 1656 (1998).
\bibitem{NA44KK}    I.~G.!Bearden {\it et al.},
 Phys. Rev. Lett. {\bf 87}, 112301 (2001).
\bibitem{StarHBT}  C. Adler {\it et al.}, Phys. Rev. Lett. {\bf  87}, 082301 (2001).
\bibitem{PhenixHBT}  K. Adcox {\it et al.}, Phys. Rev. Lett. {\bf 88}, 192302  (2002).
\bibitem{HeinzJacak}    U. W. Heinz and B. V. Jacak, Ann. Rev Nucl. Part. Sci {\bf 49} 529 (1999).
\bibitem{Akkelin95} S.~V.~Akkelin and Yu.~M.~Sinyukov, Phys. Lett.
{\bf B356}, 525 (1995).
\bibitem{Wiedemann96} U.~A.~Wiedemann, P.~Scotto, and U.~Heinz,
Phys. Rev. {\bf C53}, 918 (1996).
\bibitem{Lorstad96} T.~Cs\"{o}rg\H{o} and B.~L\"{o}rstad, Phys. Rev. {\bf C54}, 1390 (1996).
\bibitem{Adler:2002ct}
C.~Adler {\it et al.},  Phys.Rev.Lett. 90 032301 (2003).
%nucl-ex/0206006.
%``Azimuthal anisotropy and correlations in the hard scattering regime at  RHIC,''

%%CITATION = NUCL-EX 0206006;%%

\bibitem{Masashi} I.~G.~Bearden {\it et al.},  Phys. Rev. C 66, 044907 (2002).
%nucl-ex/0202019 Particle production in central Pb+Pb collisions at 158 A GeV/c 

%``Particle production in central Pb+Pb collisions at 158 A GeV/c"
\bibitem{TIC}    A. Braem {\it et al.}, Nucl. Inst. Meth. A {\bf 409} 426 (1998).
% and  {\it ibid}  C.~W. Fabjan, {\it et al.}, {\bf 367}   240 (1995).
\bibitem{NA44pp} H. B\o ggild {\it et al.}, Phys. Lett. {\bf B458} 181 (1999).
\bibitem{NA49:lambda}
S.~V.~Afanasiev {\it et al.}  %[NA49 Collaboration],
%``Lambda production in central Pb + Pb collisions at CERN-SPS energies,''
J.\ Phys.\ G {\bf 28}, 1761 (2002)
%%CITATION = NUCL-EX 0201012;%%
\bibitem{Ant00}  F. Antinori  {\it et al.}, Eur. Phys. J. {\bf C14} 633 (2000).
\bibitem{Koonin} S.~E. Koonin, Phys. Lett. {\bf 70B} 43 (1977).
\bibitem{Lednicky} R. Lednicky, V.L. Lyuboshitz, Yad. Fiz. {\bf 35} 1316 (1982) 
 (Sov. J. Nucl. Phys. {\bf 35} 770(1982)).
\bibitem{Pratt} S. Pratt, M.B. Tsang, Phys. Rev. {\bf C36} 2390 (1987).
\bibitem{Ghisal} C. Ghisalberti {\it et al.}, Nucl. Phys. {\bf A583} 401c (1995).
\bibitem{Pratt86} S.~Pratt, Phys. Rev. {\bf D33}, 72 (1986).
\bibitem{zajc84} W.~A.~Zajc {\it et al.}, Phys. Rev. {\bf C29}, 2173 (1984).
\bibitem{NA49pp}
H.~Appelshauser {\it et al.}  %[NA49 ],
%``Two-proton correlations from 158-A-GeV Pb + Pb central collisions,''
Phys.\ Lett.\ B {\bf 467}, 21 (1999).
%[arXiv:nucl-ex/9905001].
%%CITATION = NUCL-EX 9905001;%%
\bibitem{RQMD} H.~Sorge, Phys Rev C{bf 52} 3291 (1995).
%Flavor Production in Pb(160AGeV) on Pb Collisions: Effect of Color Ropes and Hadronic Rescattering,
\bibitem{Sullivan:vb} J.~P.~Sullivan {\it et al.},
%``Calculations Of Bose-Einstein Correlations From Relativistic Quantum Molecular Dynamics,''
Nucl.\ Phys.\ A {\bf 566}, 531C (1994).
\bibitem{Fields} D.~E.~Fields {\it et al.}, 
% J.~P.~Sullivan, J.~Simon-Gillo, H.~van Hecke, B.~V.~Jacak and N.~Xu,
%``Relationship Between Correlation Function Fit 
%Parameters And Source Distributions,''
Phys.\ Rev.\ C {\bf 52}, 986 (1995).

\bibitem {Nexus}
H.J.Drescher, M.Hladik, S.Ostapchenko, T.Pierog and K.Werner,
%`Parton-based Gribov-Regge theory", hep-ph/0007198.
Phys.Rev. C {\bf 65}  054902 (2002). %hep-ph/0011219 INITIAL CONDITION FOR QGP EVOLUTION FROM NEXUS.
%\bibitem{Siemiarczuk} T. Siemiarczuk, P. Zieli\'nski, Phys. Lett. {\bf 24B}   (1967) 675.
%\bibitem{Tanihata:qr} I.~Tanihata, M.~C.~Lemaire, S.~Nagamiya and S.~Schnetzer,
%``Two Proton Correlation Measurements In 800-Mev/Nucleon And 400-Mev/Nucleon Heavy Ion Reactions,''
Phys.\ Lett.\ B {\bf 97}, 363 (1980).
\bibitem{Murraydpp} M.~Murray and B. Holzer, Phys. Rev. {\bf C63} 054901 (2001).
\bibitem{SCHEIBL} R. Scheibl and U. Heinz,  Phys. Rev. C {\bf 59}, 1585 (1999).
\end{thebibliography}
\end{document}